\documentclass[10pt, reqno]{amsart}

\usepackage{amsmath,amssymb}
\usepackage{hyperref}
\usepackage{graphicx}
\usepackage{xcolor}

\usepackage{thmtools}
\usepackage{thm-restate}
\usepackage{mathrsfs}
\usepackage{todonotes}

\numberwithin{equation}{section}

\newcommand{\EE}{\mathbb{E}}
\newcommand{\RR}{\mathbb{R}}

\usepackage{relsize}
\newcommand{\T}{{\mathsmaller {\rm T}}}

\makeatletter
\newcommand*\bigcdot{\mathpalette\bigcdot@{.5}}
\newcommand*\bigcdot@[2]{\mathbin{\vcenter{\hbox{\scalebox{#2}{$\m@th#1\bullet$}}}}}
\makeatother

\begin{document}
\title[]{Treatment effect: a critique}

	\author{H.~S.~Battey}
\thanks{Department of Mathematics, Imperial College London, 180 Queen's Gate, London SW7 2AZ, UK. \newline
{\color{white}m\;} \texttt{h.battey@imperial.ac.uk} and \texttt{c.edgar24@imperial.ac.uk}}
\address{Department of Mathematics, Imperial College London}
\email{h.battey@imperial.ac.uk}

\author{Charlotte Edgar}

\begin{abstract}
Two broad positions within statistics define a treatment effect, on the one hand, as a parameter of a statistical model, and on the other, as an appropriate population-level difference in outcomes or counterfactual outcomes under the different treatment regimes. This short expository paper presents some simple but consequential insights on the two formulations, contrasting the answers under the most favourable fictitious idealisation for the counterfactual framework. These observations clarify the relationship between Fisherian model-based inference and modern counterfactual formulations, and emphasise concerns, raised by Cox and others, regarding the suitability of model-free definitions as targets of inference when scientific conclusions are intended to generalise beyond the observed sample. Parts of the paper are necessarily controversial; we follow Cox (1958a) in not putting these forward in any dogmatic spirit.

\medskip
\noindent \emph{Some key words}: causation; counterfactuals; foundations of modelling; scientific practice.
\end{abstract}

\maketitle

\section{Introduction}\label{secIntro}

When the aim is scientific understanding rather than immediate predictive success, establishing causation is usually the idealised objective even when the word is not explicitly used. The terminology \emph{causal inference} has, however, come to be primarily associated with a specific formulation based on counterfactuals, in the vein of Rubin (1974) and Rosenbaum and Rubin (1983). This has gained popularity as a basis for methodological development, particularly in connection with the 
property of double robustness first introduced by Robins \emph{et al.} (1994). Of two primary aspects that are open to discussion, we focus here on the way in which the effect of potential causes is measured, revisiting the fundamental question \emph{what is a treatment effect} (McCullagh, 2002)?

The model-free definition of treatment effect implicit in the counterfactual formulations differs strikingly from a framing in which the treatment effect is defined as a parameter of a statistical model, in the vein of Fisher (1922), who appears to have been the first to use the word \emph{parameter} in the general modern sense (Stigler, 1976). In view of this and also Bailey's (2017) correction to a historical misnomer, it is appropriate to associate with Fisher's name the model-based definition of treatment effect rather than a definition implicit in what is sometimes referred to as \emph{Fisher's sharp null hypothesis}. The model-based definition encompasses a style of discussion often suggestive of causation and reflected in regression-type formulations (e.g.~Cox, 1958b, 1968, 1972, 2007; McCullagh and Nelder, 1989).

The issue considered here is not whether model-free treatment effects based on counterfactuals can be identified or consistently estimated under suitable assumptions, but whether they define appropriate quantities in settings where a stable treatment effect is unambiguously defined. We do not claim any great originality for the points made; the merits of the work are instead to clarify the issues in an incisive form based on some idealised examples. Our position echoes concerns raised by Cox (1958b, 2012), who questioned the suitability of model-free averages as targets of inference when conclusions are intended to generalise beyond the observed sample.

\section{Model-based treatment effect}\label{secModelBased}

\subsection{Introduction}

Parametric statistical inference starts from a provisional assumption that data correspond to random variables, that these are drawn from an unknown probability distribution, and that this distribution belongs to a parametrised family $\mathcal{P}_\Theta:=\{P_\theta : \theta \in \Theta\}$, the statistical model. Dependence or independence between observations is reflected in $\mathcal{P}_\Theta$. While there is considerable flexibility in how a statistical model might be constructed, a core principle is that $\mathcal{P}_\Theta$ should stress commonalities between observational or experimental units, the purpose being to provide subject-matter understanding in such a way that any conclusions drawn remain relevant to units different from those already observed. Such commonalities are encapsulated in stable parameters, of which a treatment parameter, quantifying treatment effect, is one example.

McCullagh (2002) formalised some principles underpinning the construction of sensible statistical models, pointing out that the textbook definition does not incorporate by default aspects usually invoked, with the implication that four ``plainly absurd'' examples are in principle valid. 
His work can be viewed as an attempt to put scientific common sense on a more rigorous mathematical footing and stems to some extent from an earlier technical report (McCullagh, 1999) establishing a group theoretic perspective on generalised linear models via exponential tilting. McCullagh (2022, Ch.~11 and 14) gives an accessible exposition of some of the main ideas. An important point is that this formalisation is detached from any causal mechanism; it is rather the context that determines whether causation can be securely established.

\subsection{McCullagh's formalisation}

A treatment has an effect, not on the outcome, but on the probability distribution of the outcome. McCullagh (2022, p.~231, p.~399) expresses this as a transformation induced by the action of a group $\mathcal{G}$ on $\{P_\gamma: \gamma\in \Gamma\}$, the set of baseline distributions for the outcome prior to possible treatment. In the simplest models the group acts on the baseline distributions via the baseline parameter space $\Gamma$ as $\mathcal{G}\times \Gamma \rightarrow \Gamma$, i.e.~for any $g\in \mathcal{G}$ and $\gamma\in\Gamma$, $(g,\gamma)\rightarrow g\gamma \in \Gamma$. As two simple examples: if $\Gamma=\RR$, the natural group action is addition, $g=(\psi, +)$ say, so that $g\gamma = \gamma + \psi$; if $\Gamma=\RR^+$, the natural group action is multiplication, $g=(\psi, \times)$ say, and $g\gamma=\gamma\psi$. In each case the null treatment effect is the identity element of the group, i.e.~addition of zero in the first example and multiplication by one in the second.

In what follows, we will use the imprecise terminology \emph{individual} to refer to an experimental or observational unit of any kind. The effect of the treatment is to modify the response distribution, altering the control distribution by the same group action for all treated individuals. Implicit in this is the requirement that outcome distributions are notionally compared for individuals $i$ and $i'$, perhaps hypothetical, that are comparable in the sense that for all events $A$ in the sample space for the outcome,
\begin{align}
\begin{split}\label{eqComparable}
\text{pr}(Y_i\in A\mid T=0) =& \text{pr}(Y_{i'}\in A\mid T=0) \\
\text{pr}(Y_i \in A\mid T=1) =& \text{pr}(Y_{i'}\in A\mid T=1),
\end{split}
\end{align}
where here and henceforth we restrict attention to a binary treatment variable taking values $t\in\{0,1\}$. In the simplest parametrised models within this simplest setting, \eqref{eqComparable} implies that $i$ and $i'$ share the same baseline parameter $\gamma_i=\gamma_{i'}$, and if both are treated, both are perturbed from baseline in the same way. Dependence on intrinsic features, $w$ say, recovers standard regression formulations. Specifically, if $\Gamma=\RR$, a popular model would be the linear regression with $\gamma_i=w_i^\T \beta$, while if $\Gamma=\RR^+$, the common modelling assumption $\gamma_i=\exp(w_i^\T \beta)$ would recover some other popular regression formulations. While the group element defining the treatment effect is stable over individuals, the control distribution is  individual-specific in view of $\gamma_i$. Interaction of treatment with intrinsic variables is discussed on pp.~232-240 of McCullagh (2022), allowing some variation of treatment effects by unit while respecting the stability requirement.

\section{Model-free treatment effect and counterfactuals}

In model-free settings there is no outcome model in which to embed a natural and stable treatment parameter. For a binary treatment, a common choice is therefore an average difference in potential outcomes, the simplest model-free counterfactual treatment effect being
\begin{equation}\label{eqTau}
\tau_n := \frac{1}{n}\sum_{i=1}^{n}\EE_i(Y_{i}(1) - Y_{i}(0)),
\end{equation}
where $Y_i(t)$ is the outcome on individual $i$ for values $t\in\{0,1\}$ of the treatment indicator. See e.g.~Imbens and Rubin (2015) for a more exhaustive discussion of the framework. 
 Since, for a given $i$, only one of $Y_{i}(1)$ or $Y_{i}(0)$ is observable, two broad approaches are common, based on assumptions about the unobservable ``cross world'' of the extended counterfactual process: one is to impute the counterfactual on the basis of covariate information, requiring an outcome model; the other is propensity-score reweighting (Rosenbaum and Rubin, 1983), which models the treatment assignment mechanism instead. Under the strong assumption of no unmeasured confounding and several further assumptions, it can be shown that the resulting observable statistic, $\hat \tau_n$ say, estimates $\tau_n$. 

Although Imbens and Rubin (2015, Chapter 8) refer to model-based approaches, such models are only used as a means for performing inference on \eqref{eqTau} or a similarly defined model-free treatment effect. Their usage of this terminology is thus different, in the relevant sense, from that of \S \ref{secModelBased}.

\section{Some prompts for reflection}\label{secPrompts}

\subsection{Two aspects} 

There are two distinct types of consideration regarding these strikingly different definitions of treatment effect. One concerns the security of causal claims and the strength of untestable assumptions, and the other concerns the way in which the effect of the potential cause is measured. The first of these has been covered elsewhere in the literature, notably by Dawid (2000) and McCullagh (2022, p.~244). The criticism is that a formulation in terms of objects that are inherently unobservable can never be falsified, and therefore violates core principles of the scientific method, as put forward by e.g.~Popper (1963). We will not repeat any of this earlier discussion, focussing instead on the second consideration, which can be decoupled from the first under a fictitious idealisation for the counterfactuals.

\subsection{Fictitious idealisation}

To detach concerns over the use of counterfactuals from those over what is being measured, we suppose, notionally, that the process through which the counterfactuals become factual is completely known, and that both outcomes are observable. This fictitious idealisation is the most favourable one for the counterfactual formulation and permits isolation of key issues in an incisive form.

Under the fictitious idealisation, the counterfactual formulation reduces to a matched-pair problem in which each individual is twinned with a perfect copy of itself. From individual $i$ and his or her fictitious twin, one of the two is randomised to treatment, the other to control, yielding notionally observable outcomes $(Y_{i1}, Y_{i0})$. Given an outcome model for $(Y_{i1}, Y_{i0})$ with a stable treatment parameter, the question we seek to address is to what extent the model-free definition of treatment effect \eqref{eqTau} recovers it, and how this depends on the structure of the outcome model. If the answers are drastically different under simple idealised settings, this must raise concerns over the relevance of \eqref{eqTau} in more difficult contexts.

\subsection{Five examples}\label{secExamples}

The simplest example in which the two formulations give identical answers is a linear model of the form $Y_{i0}= \gamma_i + \varepsilon_i$ and $Y_{i1}= \gamma_i + \Delta + \varepsilon_i$ with $\varepsilon_i$ a mean-zero error term. This encompasses the linear regression model with intrinsic variables $w$ by taking $\gamma_i=w_i^\T \beta$.  Implicit in the model just presented is an assumption of unit treatment additivity that underpinned R.~A.~Fisher's early work on the design of experiments. The model-free treatment effect \eqref{eqTau} recovers the model-based treatment effect $\Delta$ in this case. This example is essentially the same as that discussed by Imbens and Rubin (2015, pp.~122).

A similarly simple example takes $Y_{i0}$ and $Y_{i1}$ as exponentially distributed of rates $\gamma_i$ and $\gamma_i\theta$ respectively, so that the model-based treatment effect is multiplicative on the rate scale. It is appropriate, since the outcomes are positive, to construct the model-free treatment effect \eqref{eqTau} after logarithmic transformation, giving $\tau_n = -\log \theta $, which is the model-based treatment effect $\theta$ converted to a different scale. The two approaches are again compatible.

The previous example can be elaborated by extending the outcome distribution to Weibull with shape parameter $\alpha$, the rate parameters being specified as above, with multiplicative treatment effect $\theta$ on the rate scale. Since the density function of the ratio $Z_i=Y_{i1}/Y_{i0}$ is
\[
f_Z(z)=\frac{\alpha \theta z^{\alpha-1}}{(1+\theta z^\alpha)^2}, \quad z\geq 0,
\]
the definition \eqref{eqTau} yields, after logarithmic transformation,
\[
\tau_n = \frac{1}{n}\sum_{i=1}^n\EE(\log Y_{i1}- \log Y_{i0}) = \int_0^\infty  \frac{\log(z) \alpha \theta z^{\alpha-1}}{(1+\theta z^\alpha)^2} dz = - \frac{\log \theta}{\alpha},
\]
showing that the true treatment effect $\theta$ can only be identified up to the sign of $\log \theta$, unless $\alpha$ is known, as in the previous exponential example with $\alpha=1$. This is an example in which the answers are semi-compatible, being qualitatively reasonable and stable across samples.

We now present two examples in which the disagreement seems more severe. Suppose that $Y_{i0}$ and $Y_{i1}$ are exponentially distributed of rates $\gamma_i$ and $\gamma_i +\Delta$ respectively. Since the density function of the ratio $Z_i=Y_{i1}/Y_{i0}$ is
\[
f_{Z_i}(z)=\frac{\gamma_i(\gamma_i + \Delta)}{(\gamma_i + (\gamma_i+\Delta)z)^2}, \quad z\geq 0,
\]
equation \eqref{eqTau} yields, after logarithmic transformation
\begin{equation}\label{eqExpAddit}
\tau_n=\frac{1}{n}\sum_{i=1}^n \log((\gamma_i+\Delta)/\gamma_i).
\end{equation}
At $\Delta=0$, the model-based and model-free treatment effects agree, otherwise $\tau_n$ depends on the sample chosen via $(\gamma_i)_{i=1}^n$.  Consider the example of $\Delta=2$ and two extreme cases: if all the $\gamma_i$ are concentrated near zero, $\tau_n$ is large, while if all the $\gamma_i$ are large, then $\tau_n$ is tiny. Thus, \eqref{eqTau} converts a well-defined problem with a stable definition of treatment effect into one that is unstable with respect to sampling new individuals, and in which extrapolation of conclusions to new individuals is impossible. Although it may be argued that deviations induced by $(\gamma_i)_{i=1}^n$ should approximately average out to a value close to $\Delta$, this seems in many contexts an unreasonable assumption. R.~A.~Bailey (2017) writes, in her critique of a paper in \emph{Statistical Science}:

\begin{quotation}
	\emph{I have practised as a statistician for 40 years} [\ldots], \emph{in no case were the experimental units a random sample from a fixed finite population. They were convenient, and were deemed to be representative enough that results on them could be extrapolated to other units.}	
\end{quotation}

One final example with a similar conclusion takes the outcomes to be binary, from the logistic model			
\[
Y_{i0}\sim\text{Bernoulli}\biggl(\frac{e^{\gamma_i}}{1+e^{\gamma_i}}\biggr), \quad  Y_{i1}\sim\text{Bernoulli}\biggl(\frac{e^{\gamma_i + 	\Delta}}{1+e^{\gamma_i + \Delta}}\biggr),
\]
logistic regression being recovered by modelling $\gamma_i=w_i^\T\beta$. The model-free treatment effect \eqref{eqTau} is
\begin{equation}\label{eqLogistic}
\tau_n=\frac{1}{n}\sum_{i=1}^{n}\frac{e^{\gamma_i}(e^\Delta - 1)}{(1+e^{\gamma_i})(e^{\gamma_i}e^\Delta + 1)}.
\end{equation}
Except at $\tau_n=\Delta=0$, there is a complicated dependence on the sample, as in \eqref{eqExpAddit}.

The model-based and model-free definitions give identical answers for any sample if and only if the following criteria are all satisfied:
\begin{enumerate}
	\item The group action defining the treatment effect acts on the baseline distribution $P_\gamma$ via the baseline parameter space, transforming it to $P_{g\gamma}$. 
	\item The group action can be described by the same group action on the observations.
	\item There are no other nuisance parameters needed to characterise the relevant expectation (on the original or logarithmic scale) under $P_\gamma$ besides $\gamma$.
\end{enumerate}
If any of these conditions is violated, \eqref{eqTau} depends in general on the values of other nuisance parameters or on the sample chosen via the baseline parameters. Conditions (1) and (2) imply that the model belongs to a transformation family (e.g.~Barndorff-Nielsen and Cox, 1994, \S 2.8).

\subsection{Clarification of Fisher's position}\label{secFisher}

The approach based on \eqref{eqTau} is sometimes called \emph{Neymannian} following Neyman (1923) and particularly Neyman \emph{et al.}~(1935). Within the causal inference community, another hypothesis of the form $Y_i(1)=Y_i(0)$ for all $i=1,\ldots,n$ is sometimes contemplated. This has come to be associated with R.~A.~Fisher's name in the form of ``Fisher's sharp null''. Bailey (2017) corrected this historical misnomer, pointing out that Fisher's formulation was instead based on unit treatment additivity on an appropriate scale, as in the first two examples of \S \ref{secExamples}, which implicitly uses the model-based definition specialised to the situation where the outcome distribution belongs to a location family, perhaps after logarithmic transformation. This also resolves Holland's (1983, \S 6) confusion regarding D.~R.~Cox's position. 
McCullagh's (2002, 2022) formalisation unifies these early contributions in the analysis of field experiments with the other examples. 

The last two examples of \S \ref{secExamples} show that the Fisherian notion of treatment effect typically violates the sharp null even under the fictitious idealisation in which the counterfactuals are observable.

\section{Differing positions}

Three arguments are commonly put forward in favour of the model-free definition of treatment effect:
\begin{itemize}
\item[A.1] That \eqref{eqTau} allows the treatment to affect different individuals in different ways. 
\item[A.2] That \eqref{eqTau} does not require an outcome model.
\item[A.3] That \eqref{eqTau} admits doubly-robust estimators.
\end{itemize}
Here we expand on each of these arguments and summarise our own position.

In connection with the so-called sharp null, defined above, Ding and Dasgupta (2016) argued that the assumption of constant causal effect across individuals is unrealistic for binary outcomes on the scale of expectations, a conclusion with which we agree in view of \eqref{eqLogistic}. This was noted in support of an argument that individual units should be allowed to contribute a different treatment effect to a composite, which is permissible under \eqref{eqTau}; this is where our positions differ. The last example of \S \ref{secExamples} indicates an appropriate way of achieving stability, in this case via a stable additive treatment effect on the log-odds scale, while allowing individuals to have different probabilities for the event of interest. A treatment effect that differs according to individual-specific features is incorporated in the model-based definition through an interaction component, which retains the necessary stability of parameters; see McCullagh (2022, pp.~232--240).

The most legitimate objection (argument A.2) of the model-based formulation is the possibility of model misspecification. By eschewing the model, stability and interpretation is lost through the use of \eqref{eqTau} or similar, but if the model is misspecified, this may also lead to a loss of stability across samples. This underscores the importance of assessments of model adequacy; see e.g.~Barndorff-Nielsen and Cox (1994, p.~29) for some discussion from a Fisherian position.

Double robustness (argument A.3) of an estimator of $\tau_n$ is the property of being consistent for $\tau_n$ either under a valid outcome model, or under a valid model for the treatment assignment, provided that other assumptions such as no unmeasured confounding are satisfied. The property of double robustness is only relevant if the definition $\tau_n$ is accepted in the first place, and it is this that we have sought to challenge in \S \ref{secPrompts}.

\section{Conclusions and reflections on a burgeoning literature}

Section \ref{secExamples} exemplifies how the model-free treatment effect \eqref{eqTau}, and other model-free definitions, are liable to change depending on the particular sample chosen, even when there is a clearly defined treatment effect that is stable over individuals. The implication is that the conclusions of inference from one study do not generalise to other populations. This observation is not new, and indeed, the ability of \eqref{eqTau} to accommodate individual-specific differences is viewed as a strength by some authors. Our view, echoing Cox (1958b, 2012), is that if there is a systematic difference in treatment effect between units, then that needs to be explained. If, in a traditional modelling setting, parameters representing treatment effects were allowed to depend in an arbitrary way on the observation index, this would be in clear violation of the foundations of modelling, as formalised by McCullagh (2002).

Instability in the definition of the model-free treatment effect \eqref{eqTau} or similar is the implicit reason behind several developments in the causal inference literature. For instance, Jin and Rothenh{\"a}usler (2024) proposed an approach to inference with conditional validity in finite populations, motivated by the observation that versions of $\tau_n$ for sub-populations are typically more relevant to individuals in those populations than a version specified as an average over a larger population. The need to introduce different values of $\tau_n$ for sub-populations is an inherent feature of their instability. In a similar vein, Chernozhukov \emph{et al.}~(2024) developed conditional influence functions to improve the efficiency of average treatment effect estimators conditional on a subset of covariates.

Other developments in the causal inference literature have emerged to address aspects that do not arise in the model-based formulation. Inverse probability weighting as discussed in Rosenbaum and Rubin (1983) is one way to correct for a lack of balance among the treatment groups prior to comparison by the model-free treatment effect \eqref{eqTau} or generalisations thereof. Other approaches have been proposed, for instance imputation of the missing counterfactual outcomes. 

In a standard regression model, there is no need for covariate classes to be balanced; maximum-likelihood inference on treatment parameters is reliable under standard regularity conditions and, while balance typically improves the precision of estimators, imbalance does not systematically bias the conclusions unless there are unmeasured confounders, a situation that is precluded by Rosenbaum and Rubin (1983) and, to our knowledge, all subsequent discussions. Moreover, since Fisherian inference is made either conditionally on the observed values of the covariates or on a lower dimensional ancillary statistic if one exists, extreme imbalance in the classes is reflected in the reference sets used for inference. An approach that averages over the covariate values observed for all individuals is, from a Fisherian perspective, too unconditional.

\bigskip

\subsection*{Funding}
	H.~Battey was supported by a UK Engineering and Physical Sciences Research Fellowship (EP/T01864X/1)

\bigskip
\bigskip

\section*{References}

\vspace{0.3cm}

Bailey, R.~A.~(2017). Inference from randomized (factorial) experiments. \emph{Statist.~Sci.}, 32, 352--355. \\

Chernozhukov, V., Newey, W.~K.~and Syrgkanis, V.~(2024). Conditional Influence Functions. \emph{arXiv: 2412.18080}. \\

Cox, D.~R.~(1958a). Some problems connected with statistical inference. \emph{Ann.~Math.~Statist.}, 29, 357--372. \\

Cox, D.~R.~(1958b). The interpretation of the effects of non-additivity in the Latin square. \emph{Biometrika}, 45, 69--73. \\

Cox, D.~R.~(2000). Causal inference without counterfactuals: comment. \emph{J.~Amer.~Statist.~Assoc.}, 95, 424--425  \\

Cox, D.~R.~(2012). Statistical causality: some historical remarks. Pages 1-5 in \emph{Causality: statistical perspectives and applications}. Edited by C.~Berzuini, P.~Dawid, L.~Bernardinelli. Wiley, Chichester. \\

Dawid, A.~P.~(2000). Causality without counterfactuals. \emph{J.~Amer.~Statist.~Assoc.}, 95, 407--424.  \\

Dawid, A.~P.~(2024). Potential outcomes and decision-theoretic foundations for statistical causality: Response to Richardson and Robins. \emph{J.~Causal Inference}, 12, paper number 20230058.  \\

Ding, P.~and Dasgupta, T.~(2016). A potential tale of two by two tables from completely randomized experiments. \emph{J.~Amer.~Statist.~Assoc.}, 111, 157--168. \\

Fisher, R.~A~(1922). On the mathematical foundations of theoretical statistics. \emph{Philos.~Trans.~Roy.~Soc.~London Ser A}, 222, 309--368. \\

Fisher, R.~A~(1935). \emph{The Design of Experiments}. Oliver and Boyd, Edinburgh.  \\

Holland, P.~(1983). Statistics and causal inference. \emph{J.~Amer.~Statist.~Assoc.}, 81, 945--960. \\

Imbens, G.~W. and Rubin, D.~B.~(2015). \emph{Causal Inference for Statistics, Social, and Biomedical Sciences: An Introduction}. Cambridge University Press, Cambridge. \\

Jin, Y.~and Rothenh{\"a}usler, D.~(2024). Tailored inference for finite populations: conditional validity and transfer across distributions. \emph{Biometrika}, 111, 215--233. \\

McCullagh, P.~and Nelder, J.~(1989). \emph{Generalized Linear Models}. Chapman and Hall, London. \\

McCullagh, P.~(1999). The algebraic structure of generalised linear models. University of Chicago, technical report number  \\

McCullagh, P.~(2002). What is a statistical model? \emph{Ann.~Statist.}, 30, 1225--1310. \\

McCullagh, P.~(2022). \emph{Ten Projects in Applied Statistics}. Springer, Cham. \\

Neyman, J., Iwaszkiewicz, K.~and Kolodziejczyk, S.~(1935). Statistical problems in agricultural experimentation (with discussion). \emph{J.~R.~Statist.~Soc.~Suppl.}, 2, 107--180. \\ 

Popper, K.~(1963). \emph{Conjectures and Refutations: The Growth of Scientific Knowledge}. Routledge, London. \\

Richardson, T.~S.~and Robins, J.~M.~(2023). Potential outcome and decision theoretic foundations for statistical causality. \emph{J.~Causal Inference}, 11, paper number 20220012. \\

Rosenbaum, P.~and Rubin, D.~B.~(1983). The central role of the propensity score in observational studies for causal effects. \emph{Biometrika}, 70, 41--55. \\

Robins, J., Rotnitzky, A.~and Zhao, L.~P.~(1994). Estimation of regression coefficients when some regressors are not always observed. \emph{J.~Amer.~Statist.~Assoc.}, 89, 846--866. \\

Rubin, D.~B.~(1974). Estimating causal effects of treatments in randomized and nonrandomized studies. \emph{J.~Educational Psychology}, 66, 688--701. \\

Stigler, S.~(1976). Discussion of `On rereading R.~A.~Fisher' by L.~J.~Savage. \emph{Ann.~Statist.}, 4, 441--500. \\

\end{document}